\documentclass[aps,prl,twocolumn,superscriptaddress,showpacs]{revtex4}

\usepackage{graphicx}

\usepackage{times}

\begin{document}


\title{Near-Threshold Photoproduction of $\Lambda(1520)$ from Protons and Deuterons}





\author{N.~Muramatsu}
  \affiliation{Research Center for Nuclear Physics, Osaka University, Ibaraki, Osaka 567-0047, Japan}
  \affiliation{Kansai Photon Science Institute, Japan Atomic Energy Agency, Kizu, Kyoto, 619-0215, Japan}
\author{J.Y.~Chen}
  \affiliation{Department of Physics, National Sun Yat-Sen University, Kaohsiung 804, Taiwan}
  \affiliation{Institute of Physics, Academia Sinica, Taipei 11529, Taiwan}
\author{W.C.~Chang}
  \affiliation{Institute of Physics, Academia Sinica, Taipei 11529, Taiwan}
\author{D.S.~Ahn}
  \affiliation{Research Center for Nuclear Physics, Osaka University, Ibaraki, Osaka 567-0047, Japan}
  \affiliation{Department of Physics, Pusan National University, Busan 609-735, Korea}
\author{J.K.~Ahn}
  \affiliation{Department of Physics, Pusan National University, Busan 609-735, Korea}
\author{H.~Akimune}
  \affiliation{Department of Physics, Konan University, Kobe, Hyogo 658-8501, Japan}
\author{Y.~Asano}
  \affiliation{Kansai Photon Science Institute, Japan Atomic Energy Agency, Kizu, Kyoto, 619-0215, Japan}
\author{S.~Dat\'{e}}
  \affiliation{Japan Synchrotron Radiation Research Institute, Sayo, Hyogo 679-5143, Japan}
\author{H.~Ejiri}
  \affiliation{Research Center for Nuclear Physics, Osaka University, Ibaraki, Osaka 567-0047, Japan}
  \affiliation{Japan Synchrotron Radiation Research Institute, Sayo, Hyogo 679-5143, Japan}
\author{H.~Fujimura}
  \affiliation{Department of Physics, Kyoto University, Kyoto 606-8502, Japan}
  \affiliation{Laboratory of Nuclear Science, Tohoku University, Sendai, Miyagi 982-0826, Japan}
\author{M.~Fujiwara}
  \affiliation{Research Center for Nuclear Physics, Osaka University, Ibaraki, Osaka 567-0047, Japan}
  \affiliation{Kansai Photon Science Institute, Japan Atomic Energy Agency, Kizu, Kyoto, 619-0215, Japan}
\author{S.~Fukui}
  \affiliation{Department of Physics, Nagoya University, Aichi 464-8602, Japan}
\author{S.~Hasegawa}
  \affiliation{Research Center for Nuclear Physics, Osaka University, Ibaraki, Osaka 567-0047, Japan}
\author{K.~Hicks}
  \affiliation{Department of Physics and Astronomy, Ohio University, Athens, Ohio 45701, USA}
\author{K.~Horie}
  \affiliation{Research Center for Nuclear Physics, Osaka University, Ibaraki, Osaka 567-0047, Japan}
\author{T.~Hotta}
  \affiliation{Research Center for Nuclear Physics, Osaka University, Ibaraki, Osaka 567-0047, Japan}
\author{K.~Imai}
  \affiliation{Department of Physics, Kyoto University, Kyoto 606-8502, Japan}
\author{T.~Ishikawa}
  \affiliation{Laboratory of Nuclear Science, Tohoku University, Sendai, Miyagi 982-0826, Japan}
\author{T.~Iwata}
  \affiliation{Department of Physics, Yamagata University, Yamagata 990-8560, Japan}
\author{Y.~Kato}
  \affiliation{Research Center for Nuclear Physics, Osaka University, Ibaraki, Osaka 567-0047, Japan}
\author{H.~Kawai}
  \affiliation{Department of Physics, Chiba University, Chiba 263-8522, Japan}
\author{K.~Kino}
  \affiliation{Research Center for Nuclear Physics, Osaka University, Ibaraki, Osaka 567-0047, Japan}
\author{H.~Kohri}
  \affiliation{Research Center for Nuclear Physics, Osaka University, Ibaraki, Osaka 567-0047, Japan}
\author{N.~Kumagai}
  \affiliation{Japan Synchrotron Radiation Research Institute, Sayo, Hyogo 679-5143, Japan}
\author{S.~Makino}
  \affiliation{Wakayama Medical University, Wakayama, 641-8509, Japan}
\author{T.~Matsuda}
  \affiliation{Department of Applied Physics, Miyazaki University, Miyazaki 889-2192, Japan}
\author{T.~Matsumura}
  \affiliation{National Defence Academy in Japan, Yokosuka, Kanagawa 239-8686, Japan}
\author{N.~Matsuoka}
  \affiliation{Research Center for Nuclear Physics, Osaka University, Ibaraki, Osaka 567-0047, Japan}
\author{T.~Mibe}
  \affiliation{Department of Physics and Astronomy, Ohio University, Athens, Ohio 45701, USA}
\author{M.~Miyabe}
  \affiliation{Department of Physics, Kyoto University, Kyoto 606-8502, Japan}
\author{M.~Miyachi}
  \affiliation{Department of Physics, Tokyo Institute of Technology, Tokyo, 152-8551, Japan}
\author{T.~Nakano}
  \affiliation{Research Center for Nuclear Physics, Osaka University, Ibaraki, Osaka 567-0047, Japan}
\author{M.~Niiyama}
  \affiliation{Department of Physics, Kyoto University, Kyoto 606-8502, Japan}
  \affiliation{RIKEN, The Institute of Physical and Chemical Research, Wako, Saitama 351-0198, Japan}
\author{M.~Nomachi}
  \affiliation{Department of Physics, Osaka University, Toyonaka, Osaka 560-0043, Japan}
\author{Y.~Ohashi}
  \affiliation{Japan Synchrotron Radiation Research Institute, Sayo, Hyogo 679-5143, Japan}
\author{H.~Ohkuma}
  \affiliation{Japan Synchrotron Radiation Research Institute, Sayo, Hyogo 679-5143, Japan}
\author{T.~Ooba}
  \affiliation{Department of Physics, Chiba University, Chiba 263-8522, Japan}
\author{D.S.~Oshuev}
  \affiliation{Institute of Physics, Academia Sinica, Taipei 11529, Taiwan}
\author{C.~Rangacharyulu}
  \affiliation{Department of Physics and Engineering Physics, University of Saskatchewan, Saskatoon SK S7N 5E2, Canada}
\author{A.~Sakaguchi}
  \affiliation{Department of Physics, Osaka University, Toyonaka, Osaka 560-0043, Japan}
\author{P.M.~Shagin}
  \affiliation{School of Physics and Astronomy, University of Minnesota, Minneapolis, Minnesota 55455, USA}
\author{Y.~Shiino}
  \affiliation{Department of Physics, Chiba University, Chiba 263-8522, Japan}
\author{H.~Shimizu}
  \affiliation{Laboratory of Nuclear Science, Tohoku University, Sendai, Miyagi 982-0826, Japan}
\author{Y.~Sugaya}
  \affiliation{Department of Physics, Osaka University, Toyonaka, Osaka 560-0043, Japan}
\author{M.~Sumihama}
  \affiliation{Research Center for Nuclear Physics, Osaka University, Ibaraki, Osaka 567-0047, Japan}
\author{Y.~Toi}
  \affiliation{Department of Applied Physics, Miyazaki University, Miyazaki 889-2192, Japan}
\author{H.~Toyokawa}
  \affiliation{Japan Synchrotron Radiation Research Institute, Sayo, Hyogo 679-5143, Japan}
\author{A.~Wakai}
  \affiliation{Akita Research Institute of Brain and Blood Vessels, Akita, 010-0874, Japan}
\author{C.W.~Wang}
  \affiliation{Institute of Physics, Academia Sinica, Taipei 11529, Taiwan}
\author{S.C.~Wang}
  \affiliation{Institute of Physics, Academia Sinica, Taipei 11529, Taiwan}
\author{K.~Yonehara}
  \affiliation{Department of Physics, Konan University, Kobe, Hyogo 658-8501, Japan}
\author{T.~Yorita}
  \affiliation{Research Center for Nuclear Physics, Osaka University, Ibaraki, Osaka 567-0047, Japan}
  \affiliation{Japan Synchrotron Radiation Research Institute, Sayo, Hyogo 679-5143, Japan}
\author{M.~Yoshimura}
  \affiliation{Institute for Protein Research, Osaka University, Suita, Osaka, 565-0871, Japan}
\author{M.~Yosoi}
  \affiliation{Research Center for Nuclear Physics, Osaka University, Ibaraki, Osaka 567-0047, Japan}
  \affiliation{Department of Physics, Kyoto University, Kyoto 606-8502, Japan}
\author{R.G.T.~Zegers}
  \affiliation{National Superconducting Cyclotron Laboratory, Michigan State University, East Lansing, Michigan 48824-1321, USA}
\collaboration{LEPS Collaboration}
\noaffiliation

\date{\today}

\begin{abstract}
  Photoproduction of $\Lambda$(1520) with liquid hydrogen and deuterium targets was examined 
at photon energies below 2.4 GeV in the SPring-8 LEPS experiment. For the first time, the differential 
cross sections were measured at low energies and with a deuterium target. A large asymmetry of 
the production cross sections from protons and neutrons was observed at backward K$^{+/0}$ angles. 
This suggests the importance of the contact term, which coexists with t-channel K exchange under
gauge invariance. This interpretation was compatible with the differential cross sections, 
decay asymmetry, and photon beam asymmetry measured in the production from protons at forward K$^+$ angles.
\end{abstract}


\maketitle

  Recently the $\Lambda$(1520) hyperon has received a lot of attention since its mass 
is close to that of the claimed pentaquark $\Theta^+$ \cite{pdg} with an opposite 
strangeness. Because of these features, photoproduction of $\Lambda$(1520) from 
protons bears a close resemblance to that of $\Theta^+$ from neutrons. While 
the experimental results of $\Lambda$(1520) photoproduction are available in the 
photon energy range of 2.8$-$4.8 GeV from the LAMP2 Collaboration \cite{lamp2}, 
this reaction has not been well understood near the threshold. Some theoretical 
calculations of the production cross sections were performed by considering a large
contribution from t-channel vector kaon (K$^*$) exchange based on the cross sections 
and decay asymmetry measured by LAMP2 \cite{titov,sibirt}. However, the LAMP2 results
could also be described by a model emphasizing the importance of 
a contact term \cite{lam1520}, which was 
necessary to conserve gauge invariance along with t-channel pseudoscalar 
kaon (K) exchange \cite{qfth}. In this framework, a strong asymmetry 
is distinctively predicted in the $\Lambda$(1520) photoproduction cross sections 
from protons and neutrons because 
a dominant contribution from the contact term is absent in the production from neutrons 
\cite{lam1520,karliner}. 
Nevertheless, the cross section with a neutron target was not available. At low energies, 
the contribution from the K$^*$ exchange was suggested to be small in a chiral unitary 
model \cite{kexch}. Indeed, no dominance of the K$^*$ exchange was observed in the decay 
asymmetry measured by the CLAS Collaboration in $\Lambda$(1520) electroproduction at
total c.m.~energies up to 2.65 GeV \cite{clas-el}. It is unclear whether energy 
or photon-virtuality dependences accounts for the difference of the decay asymmetries in 
LAMP2 and CLAS. In most of theoretical models, the cross sections at E$_\gamma$$\sim$2
GeV are predicted to be larger than those measured by LAMP2, although the difference in 
the model predictions is not small.

  In this Letter, we report the measurements of differential cross sections, decay 
asymmetry, and photon beam asymmetry of $\Lambda$(1520) photoproduction from protons
and deuterons at photon energies below 2.4 GeV. We detected two charged tracks in 
the final state of K$^+$K$^-$p from protons or K$^0$K$^-$p from neutrons using a forward
spectrometer, and the $\Lambda$(1520) production was identified by invariant and missing 
mass techniques. Backward and forward productions of $\Lambda$(1520) in the c.m.~system 
were examined by detecting a K$^+$K$^-$ or K$^+$p pair and a K$^-$p pair, respectively.
These three detection modes were complementary in the acceptance. Results in all the
detection modes will be presented for runs with a hydrogen target, while only those 
in the K$^-$p detection mode will be shown for runs with a deuterium target.

  The experiment was carried out in 2002$-$2003 at the SPring-8 LEPS facility using 
a linearly polarized photon beam produced by backward Compton scattering of Ar laser 
light from 8 GeV electrons. Photons in the energy range of 1.5$-$2.4 GeV were tagged 
by detecting the recoil electrons. The photon beam with an intensity of $\sim$10$^6$ 
/sec was alternatively injected into liquid hydrogen or deuterium targets inside 
a 15 cm-thick cell. The direction of linear polarization was controlled vertically 
or horizontally by using a half-wave plate for the laser with a polarization of 
nearly 100\%. Charged particles were detected to analyze their momenta by the LEPS
forward spectrometer \cite{kplus}, which covered $\pm$20$^\circ$ and $\pm$10$^\circ$ in the 
horizontal and vertical directions, respectively. Time of flight from the target to 
a plastic scintillator wall 4 meters downstream was measured for particle identification. 
Details of the experimental setup can be found in 
Ref.~\cite{kplus}. The integrated number of tagged photons reached 2.8$\times$10$^{12}$ 
(4.6$\times$10$^{12}$) for the hydrogen (deuterium) runs.

\begin{figure}[b]
\includegraphics{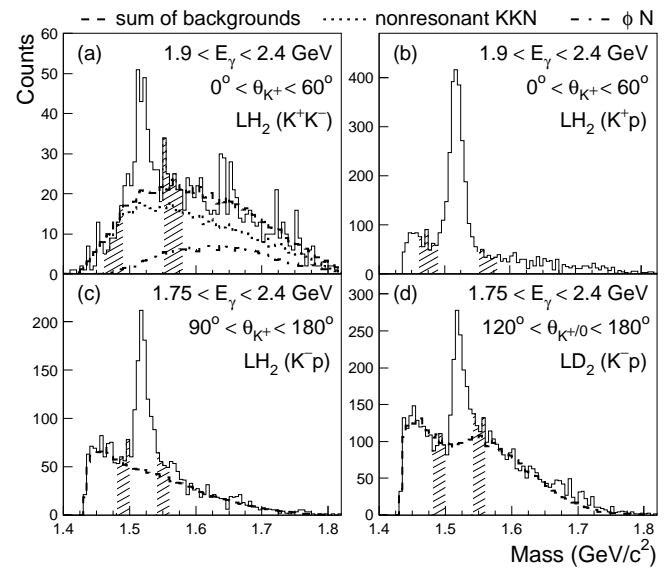}%
\caption{Mass spectra in the four analyzed samples. Panels (a) and (b) show K$^+$ missing mass spectra 
         for the hydrogen runs in the K$^+$K$^-$ and K$^+$p detection modes, respectively. Panels (c) 
         and (d) show K$^-$p invariant mass spectra in the K$^-$p detection mode for the hydrogen 
         and deuterium runs, respectively. Background spectra based on Monte Carlo simulations 
         are overlaid, while typical sideband definitions are indicated by the hatched area. 
         The backgrounds estimated in (c) and (d) are mostly due to nonresonant K$\bar{\rm K}$p 
         photoproduction. \label{f:mkp_spectra}}
\end{figure}
  The charged particles were identified within 3$\sigma$ or 4$\sigma$ of the momentum-dependent 
mass resolution. A particle decaying in flight was removed by requiring good track fitting 
qualities. The vertex point of two tracks was required to be within the target volume and the 
beam size. The missing mass of K$^+$K$^-$, K$^+$p, or K$^-$p from the proton was required to be 
around the proton, K$^-$, or K$^+$ mass \cite{pdg}, respectively. Minimum photon energies of 
1.90, 1.90, and 1.75 GeV were required in the analyses with K$^+$K$^-$, K$^+$p, and K$^-$p 
detections, respectively, due to the limitation of the acceptance for $\Lambda$(1520) 
photoproduction. Background from $\phi$ photoproduction was removed in the K$^+$K$^-$ (K$^+$p) 
detection mode by requiring the K$^+$K$^-$ invariant mass (proton missing mass) to be greater 
than 1.030 (1.050) GeV/c$^2$. In the K$^-$p detection mode, such a condition was not introduced 
because $\phi$ events were kinematically limited. A resonance peak of $\Lambda$(1520) was 
identified by K$^+$ missing mass from the proton in the K$^+$K$^-$ and K$^+$p detection modes, 
and by K$^-$p invariant mass in the K$^-$p mode, as shown in Fig.~\ref{f:mkp_spectra}. 
The reconstructed peak positions and widths are consistent with the nominal values (M$=$1519.5 
MeV/c$^2$ and $\Gamma$$=$15.6 MeV/c$^2$) \cite{pdg} convoluted with the experimental mass 
resolutions ($\sigma_M$), where the contributions from momentum and photon energy resolutions 
are 2 MeV/c$^2$ and 8 MeV/c$^2$, respectively. Statistics 
in the K$^+$K$^-$ mode were relatively low, so that the K$^+$ polar angle region examined for
the measurements of cross sections and asymmetries was limited up to 60$^\circ$ from the incident
photon direction in the c.m.~system. 
Instead, a larger angular region up to 90$^\circ$ was explored in the K$^+$p mode. In the K$^-$p 
mode, a K$^+$ (K$^{+/0}$) polar angle region above 90$^\circ$ (120$^\circ$), defined in the c.m.~system
of photon and target nucleon, was examined for the hydrogen (deuterium) 
runs. The smaller acceptance for the deuterium runs was caused by an additional requirement that 
the K$^-$p missing mass assuming a target mass of the deuteron was smaller than 1.51 GeV/c$^2$. 
This condition was introduced to avoid possible contaminations from additional reactions including 
$\gamma$d$\to$$\Lambda$(1520)$\Theta^+$ \cite{leps-lth}. The cross sections remained unchanged 
within statistical errors even by removing this condition.

  In the K$^+$K$^-$ and K$^-$p detection modes, the background level under the $\Lambda$(1520) 
resonance was estimated by two independent methods. The first method was developed on the basis of
Monte Carlo (MC) simulations, which produced background spectra for the two major photoproduction 
processes of $\phi$p and nonresonant K$\bar{\rm K}$p final states. Small contributions other than 
these processes and $\Lambda$(1520) production were included in the category of ``nonresonant'' 
process. The two background processes were generated by assuming a constant matrix element in 
a {\footnotesize GEANT3}-based \cite{geant3} MC simulation package. In addition, the K$\Lambda$(1520) 
photoproduction was simulated by varying the width of the resonance depending on the phase space 
of decay products with the Blatt-Weisskopf barrier penetration model \cite{bwbpf}. For the simulated 
events, the distributions of polar angles and momenta of the detected tracks and c.m.~energy were 
adjusted by skimming events to reproduce the real distributions in the hydrogen runs. 
The skimmed MC samples were normalized so that the sum 
of all the samples should match the real spectra of invariant and missing masses. Quasifree 
photoproduction spectra from deuterons were estimated just by adopting influence of Fermi motion 
in the above MC samples. An off-shell correction was taken into account 
for a nucleon target inside deuterium. 
Total background spectra were expressed by summing the $\phi$p and nonresonant 
K$\bar{\rm K}$p processes, as overlaid in Fig.~\ref{f:mkp_spectra}. The number of $\Lambda$(1520) 
signals was counted in 1.49$-$1.55 GeV/c$^2$ (1.50$-$1.54 GeV/c$^2$) over the background estimate 
in the K$^+$K$^-$ (K$^-$p) mode. 

  In the second method, called ``sideband subtraction'', 
the background level was estimated based on two sideband regions adjacent 
to the $\Lambda$(1520) signal window, as shown by the hatched area in Fig.~\ref{f:mkp_spectra}. 
The average of yields in the sidebands, which were of the same mass width as the signal window, 
was used as the background contribution under the resonance peak. The nonlinearity of the background 
distribution was corrected based on the overall MC spectra of $\phi$p and nonresonant K$\bar{\rm K}$p 
productions. This correction was significant in the K$^+$K$^-$ mode, while it was negligible in 
the K$^-$p mode. Overestimation of background due to the $\Lambda$(1520) tails in the sidebands 
was also corrected by using K$\Lambda$(1520) MC events.

  Since many possible background processes other than the K$^+$K$^-$p final state were 
involved in the K$^+$p detection mode, two-step sideband subtractions were adopted for
the background estimation instead of performing MC simulations. First, the K$^+$ missing 
mass spectrum was constructed for the sample which was selected within the K$^-$ mass 
window of K$^+$p missing mass ($\pm$25 MeV/c$^2$). In order to ensure only the contribution 
from the K$^+$K$^-$p final state, this spectrum was corrected by subtracting the K$^+$ 
missing mass spectrum for the sample within the sideband regions of the K$^-$ mass window.
Further sideband subtraction was 
performed in the corrected K$^+$ missing mass spectrum by setting the $\Lambda$(1520) 
signal window to 1.49$-$1.55 GeV/c$^2$, as shown in Fig.~\ref{f:mkp_spectra}(b). The linearity 
of the background spectra was reasonable in these sideband subtractions because 
only 5\% variation was observed in the results with a change of the signal window widths.

\begin{figure}[b]
\includegraphics{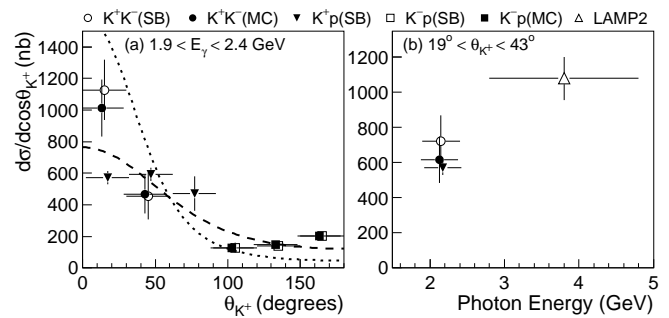}%
\caption{(a) Differential cross sections from protons at 1.9$<$ E$_\gamma$$<$2.4 GeV. The data 
         points at forward K$^+$ angles come from the K$^+$K$^-$ and K$^+$p detection modes,
         while those at backward angles are from the K$^-$p mode. The measurements using
         sideband-subtractions (SB) and Monte Carlo (MC) simulations are simultaneously plotted. 
         The dashed line (dotted line) indicates 
         a theoretical prediction for E$_\gamma$$=$1.85 (2.35) GeV with K exchange and a contact 
         term \cite{nampriv}. (b) Differential cross sections from this work and LAMP2 at 
         19$^\circ$$<$$\theta_{K^+}$$<$43$^\circ$. \label{f:dxsec_kk}}
\end{figure}
  Differential cross sections were measured in 30$^\circ$ bins in K$^{+/0}$ polar angle 
($\theta_{K^{+/0}}$) in the c.m.~system of photon and target nucleon. 
Acceptance factors were estimated in MC simulation 
taking into account the measured decay asymmetry of $\Lambda$(1520). We confirmed that the 
ratio of luminosities estimated for the hydrogen and deuterium runs was consistent with that 
of $\Lambda$(1520) signal counts in the two data sets with detection of all three tracks in the
K$^+$K$^-$p final state, which only arose from the interaction 
with protons. Figure~\ref{f:dxsec_kk}(a) shows differential cross sections from protons at 
1.9$<$E$_\gamma$$<$2.4 GeV in all the detection modes. Cross sections at forward K$^+$ angles are 
more than 3 times larger than those at backward angles. This tendency does not contradict 
the existing theoretical pictures \cite{titov,sibirt,lam1520,kexch} as Figure~\ref{f:dxsec_kk}(a)
shows a comparison with the prediction from a model, which considers the dominance of a contact-term
contribution but no contribution from the K$^*$ exchange under a rescaling by 
a cutoff mass of 650 MeV \cite{lam1520,nampriv}.
In the region of 0$^\circ$$<$$\theta_{K^+}$$<$30$^\circ$, we have not resolved the discrepancy of results
obtained in the two detection modes, but this may be caused by an interference with
different background compositions. In Fig.~\ref{f:dxsec_kk}(b), the differential cross sections 
measured at 19$^\circ$$<$$\theta_{K^+}$$<$43$^\circ$ are compared with the LAMP2 result. 
Theoretical calculations predict $\sim$1 $\mu$b or more at E$_\gamma$$\sim$2 GeV with inputs 
from the LAMP2 result \cite{titov,sibirt,lam1520,kexch}, and the present results are smaller 
than those predictions. This measurement shall provide new information for theoretical models. 

\begin{figure}[b]
\includegraphics{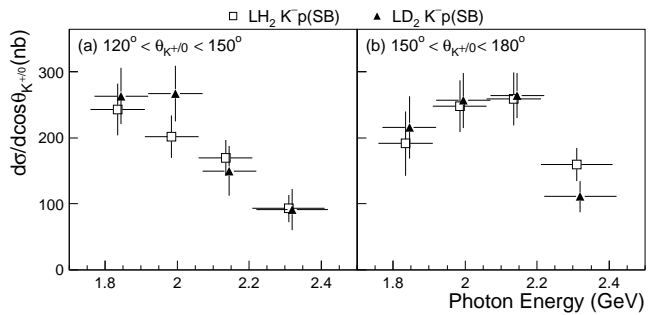}%
\caption{Differential cross sections at backward K$^{+/0}$ angles in the K$^-$p detection mode.
         Results from the hydrogen and deuterium runs are simultaneously plotted as a function 
         of photon energy for (a) 120$^\circ$$<$$\theta_{K^{+/0}}$$<$150$^\circ$ and (b) 
         150$^\circ$$<$$\theta_{K^{+/0}}$$<$180$^\circ$.\label{f:dxsec_kp}}
\end{figure}

In the K$^-$p detection 
mode, the differential cross sections from protons and deuterons were compared as shown in 
Fig.~\ref{f:dxsec_kp}. Here only the estimations using sideband subtractions are shown, 
and their deviations from 
MC-based results are typically 7\%. By combining the cross sections measured in the range of
1.75$<$E$_\gamma$$<$2.4 GeV and 120$^\circ$$<$$\theta_{K^{+/0}}$$<$180$^\circ$, the ratio of 
production from deuterons to that from protons was measured to be 1.02$\pm$0.11 in the 
sideband-subtraction method. The $\Lambda$(1520) photoproduction from neutrons was found 
to be strongly suppressed 
at backward K$^0$ angles. The observation of a large asymmetry between the productions from 
protons and neutrons conflicts with the model considering a dominance of t-channel K$^*$ exchange 
\cite{titov}, but can be explained by the model where the contact term plays a major role \cite{lam1520}. 
Larger cross sections were observed at lower energies as shown in Fig.~\ref{f:dxsec_kp}, and 
this behavior is qualitatively consistent with a theoretical calculation in Ref.~\cite{lam1520}. In the 
backward K$^+$ production from protons, cross sections also show an increase toward $\theta_{K^+}$$=$180$^\circ$
as shown in Fig.~\ref{f:dxsec_kk}(a). This may indicate an additional contribution from u-channel 
diagrams, which are conventionally considered to be small in the theoretical models 
\cite{titov,sibirt,lam1520,kexch}.

\begin{figure}[b]
\includegraphics{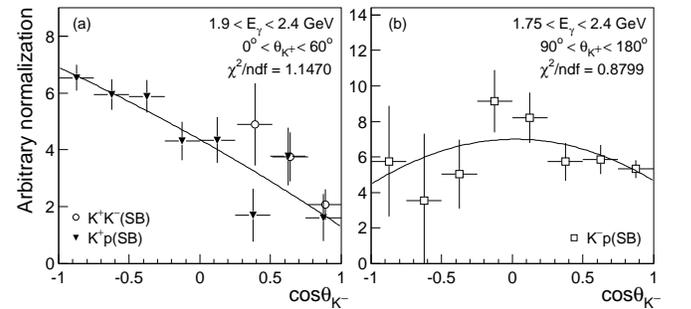}%
\caption{K$^-$ polar angle distributions in the t-channel helicity frame of $\Lambda$(1520),
         produced from protons. Panels (a) and (b) show the results at forward and backward K$^+$ angles, 
         respectively. The solid lines indicate fits by a linear combination of two spin-projection
         terms and an interference term. \label{f:thel_raw}}
\end{figure}

  The decay asymmetry of $\Lambda$(1520)$\to$K$^-$p was studied in the production from protons 
by examining the distribution of K$^-$ polar angle ($\theta_{K^-}$) in the t-channel helicity frame 
(Gottfried-Jackson frame) \cite{gjfr}. In the case of K$^*$ exchange, where the spin projection 
of helicity state is mostly of S$_z$$=$$\pm$3/2, the K$^-$ angular distribution is characterized 
by $\sin^2\theta_{K^-}$, while K exchange, composed of only S$_z$$=$$\pm$1/2, results in an angular 
distribution of $\frac{1}{3}+\cos^2\theta_{K^-}$. At the same time, a contact-term contribution, 
whose dominance is favored by the cross section measurements, is shown to possess a large component 
of S$_z$$=$$\pm$3/2 \cite{lam1520}, and the relative strength of the contact term to the K exchange 
is determined by the KN$\Lambda$(1520) coupling constant under gauge invariance \cite{lam1520,qfth}.
Figure~\ref{f:thel_raw}(a) 
shows the K$^-$ angular distribution using the sideband-subtraction methods in the K$^+$p and K$^+$K$^-$ 
detection modes, which cover forward K$^+$ angles. Although an experimental acceptance in the 
K$^+$K$^-$ mode is limited to the forward K$^-$ direction, the K$^-$ angular distribution in this 
mode is consistent with that in the K$^+$p mode. The obviously asymmetric distribution suggests 
a large interference, which is stronger than the results in LAMP2 \cite{lamp2} and CLAS \cite{clas-el}. 
We performed a fit to the K$^-$ angular distribution measured in the K$^+$p mode by using a function
$N (\alpha \sin^2\theta_{K^-} + (1-\alpha) (\frac{1}{3}+\cos^2\theta_{K^-}) + \beta \cos\theta_{K^-})$
\cite{clas-el}, where the last term represents an interference with spin-1/2 background hyperons. 
The fraction of S$_z$$=$$\pm$3/2
component ($\alpha$) was measured to be 0.520$\pm$0.063 in the photon energy range of 1.9$-$2.4 GeV, 
and clearly differed from the measurement by LAMP2 at higher energies, where the corresponding fraction
was evaluated to be 0.880$\pm$0.076 by a fit to Fig.~3 of Ref.~\cite{lamp2}. The present result was 
closer to the value from the low-energy electroproduction by CLAS \cite{clas-el}, which obtained 
0.446$\pm$0.038 for the virtual photon invariant mass of 0.9$<$Q$^2$$<$1.2 GeV$^2$. One possible way 
to accommodate the observed ratio of the two spin projections under the dominance of a contact-term 
contribution is to introduce a small contribution from the K$^*$ exchange with a destructive interference 
between S$_z$$=$$\pm$3/2 components, as shown in Fig.~13 of Ref.~\cite{lam1520}. The K$^*$N$\Lambda$(1520) 
coupling constant, which controls the strength of such an interference, is experimentally undetermined,
and the theoretical predictions of its absolute value vary in the range of 0$-$15 \cite{kstarcc}. 
The same fitting procedure was also performed for the K$^-$ angular distribution at backward K$^+$ 
angles in the K$^-$p detection mode. Figure~\ref{f:thel_raw}(b) shows the result with the sideband 
subtraction. While the interference was found to be weak, the fraction of S$_z$$=$$\pm$3/2 was 
measured to be 0.631$\pm$0.106 (0.545$\pm$0.076) in the sideband-subtraction (MC-based) method for the photon 
energy range of 1.75$-$2.4 GeV. These fractions were similar to the result in the forward 
K$^+$ direction.

  The photon beam asymmetry was also measured in the production from protons with K$^+$p 
detection for the K$^+$ angles less than 60$^\circ$. $\Lambda$(1520) signals produced from 
vertically and horizontally polarized photons were separately counted (N$_V$ and N$_H$,
respectively) as a function of K$^+$ azimuthal angle ($\phi$). The beam asymmetry ($\Sigma$) 
was measured by the fit based on an equation $\Sigma$P$_\gamma$$\cos{2\phi}$ $=$ 
(kN$_V$$-$N$_H$)/(kN$_V$$+$N$_H$) \cite{kplus}, where P$_\gamma$ was the polarization 
of the photon beam and k was a normalization factor of the two polarization samples. 
We determined $\Sigma$ to be $-$0.01$\pm$0.07, which suggests zero within uncertainties. This value is 
consistent with the prediction of the theoretical model where the contact-term contribution is dominant 
\cite{bmasy}, and thus the contribution from the t-channel K$^*$ exchange is suggested to 
be small. 

  In summary, we studied $\Lambda$(1520) photoproduction with liquid hydrogen and deuterium 
targets at E$_\gamma$$=$1.75$-$2.4 GeV. At backward K$^{+/0}$ angles, we compared the cross 
sections from protons and deuterons. A strong suppression of the production from neutrons was observed 
as suggested in the theory advocating the importance of the contact term. 
By the decay asymmetry in
the production from protons, the contribution from the S$_z$$=$$\pm$3/2 component was 
determined to be nearly 50\%. This fact may be explained by postulating a destructive
interference between the S$_z$$=$$\pm$3/2 components. The forward 
enhancement of the differential cross section and the small value of the photon beam asymmetry 
are also compatible with the interpretation adopting the contact term. 
Future precise measurements of the decay asymmetry and the photon beam asymmetry shall provide 
more accurate constraints on the value of K$^*$N$\Lambda$(1520) coupling constant as well as 
details of reaction dynamics. Further theoretical studies will also be desired at a more
quantitative level to explain the two asymmetry measurements.

  This work is closely related to the possible production of $\Theta^+$. First, the LEPS experiment 
has reported the possibility of coherent photoproduction of $\gamma$d$\to$$\Lambda$(1520)$\Theta^+$ 
\cite{leps-lth}, where forward $\Lambda$(1520) production plays an important role to induce the 
reaction of K$^+$n$\to$$\Theta^+$. The present measurement of elementary cross sections is essential to 
advance theoretical calculations of this reaction. Secondly, a comparison of the results in the 
LEPS \cite{leps-qf} and CLAS \cite{clasp2} experiments might hint that $\Theta^+$ photoproduction 
from protons is suppressed relative to that from neutrons. Such a target-isospin asymmetry may 
originate from the contact term \cite{qfth} as suggested by the forward $\Lambda$(1520) 
photoproduction in the present work.


\begin{acknowledgments}
The authors thank A. Hosaka and S.-i. Nam for theoretical discussions. We gratefully acknowledge 
the support of the staff at SPring-8 for providing excellent experimental conditions. This 
research was supported in part by the Ministry of Education, Science, Sports and Culture of 
Japan, the National Science Council of Republic of China (Taiwan), and the KOSEF of Republic 
of Korea.
\end{acknowledgments}


\end{document}